\documentclass[prl,aps,twocolumn,showpacs]{revtex4}

\usepackage{graphicx}

\begin{document}

  \title{Origin of the Structural and
    Magnetic Anomaly of the Layered Compound SrFeO$_2$:\\
    A Density
    Functional Investigation}

  \author{H. J. Xiang}
  \affiliation{National Renewable Energy Laboratory, Golden, Colorado
    80401}

  \author{Su-Huai Wei}

  \affiliation{National Renewable Energy Laboratory, Golden, Colorado 80401}

  \author{M.-H. Whangbo}

  \affiliation{Department of Chemistry, North Carolina State
    University, Raleigh, North Carolina 27695-8204}

  \date{\today}

  \begin{abstract}
 The structural and magnetic anomaly of the 
    layered compound SrFeO$_{2}$
    were examined by first principles 
    density functional calculations
    and Monte Carlo simulations.
    The down-spin Fe 3$d$ electron occupies the
    $d_{z^2}$ level rather than the degenerate ($d_{xz}$,
    $d_{yz}$) levels,
    which explains the absence of Jahn-Teller instability,
    the easy ab-plane magnetic anisotropy 
    and the observed three-dimensional (0.5, 0.5, 0.5)
    antiferromagnetic
    order.
    Monte Carlo simulations show that the strong
    inter-layer spin exchange is essential for the high N\'eel
    temperature. 
  \end{abstract}


  \maketitle
  Perovskite oxides have attracted considerable
  interest due to their extensive applications in a number of
  technological areas.
  Among them, SrFeO$_{3-x}$ and its related iron perovskite oxides exhibit
  fast oxygen transport and high electron conductivity even at low
  temperatures \cite{Shao2004,Falcon2002,Badwal2001,Wang2001,Lebon2004}.
  It had previously been believed that the
  end member phases for these compounds are the orthorhombic
  brownmillerite SrFeO$_{2.5}$ ($x=0.5$) and
  the cubic
  perovskite SrFeO$_{3}$ ($x=0$). Very recently, the range of $x$ was
  extended to $x=1$ by Tsujimoto {\it et al.}, who discovered that
  SrFeO$_{2}$ has planar FeO$_{2}$ layers made up of
  corner-sharing FeO$_{4}$ squares with high-spin Fe$^{2+}$
  ($d^{6}$) ions, separated by Sr$^{2+}$ ions \cite{Tsujimoto2007,Hayward2007}.

  SrFeO$_{2}$ exhibits
  interesting and  apparently puzzling physical
  properties \cite{Tsujimoto2007}. First,
  if the lone down-spin
  electron of a high-spin Fe$^{2+}$ ($d^{6}$) ion at square-planar site occupies the degenerate
  ($d_{xz}$,$d_{yz}$) orbitals, as expected by the crystal field theory
  \cite{Wells1962}, SrFeO$_2$ should be subject to orbital ordering or
  Jahn-Teller distortion when the temperature is lowered
  \cite{Murakami1998}. However, SrFeO$_{2}$ shows no structural
  instability and maintains the space group $P4/mmm$ down to 4.2 K
  \cite{Tsujimoto2007}.
  Second,
  SrFeO$_{2}$ displays a three-dimensional (3D) antiferromagnetic (AFM)
  order with a very high N\'eel
  temperature ($T_N = 473$ K) \cite{Tsujimoto2007}, which is even
  higher than that ($\sim 200$ K) of FeO with a 3D structure. Such a
  high 3D AFM ordering temperature in a layered system
  is remarkable and unexpected, because $T_N$ usually decreases
  drastically
  when the dimensionality decreases. Third, the powder neutron
  diffraction
  study shows that the magnetic moments are perpendicular to the
  $c$-axis (the local $z$-axis) \cite{Tsujimoto2007}, which is not
  consistent
  with the occupation of the ($d_{xz}$,$d_{yz}$) orbitals
  with three electrons
  \cite{Dai2005}.

  \begin{table}
    \caption{Spin exchange parameters (in meV) from the LDA$+$U
      calculations. The spin exchange paths $J_1$, $J_2$, $J_3$, and $J_4$ are defined in
      Fig.~\ref{fig1}. Positive (negative) values indicate that the spin
      exchange  interactions are AFM (FM).
      For SrFeO$_2$, both the experimental (expt.) \cite{Tsujimoto2007}
      and the optimized (opt.) crystal structures were used for the
      calculations.
      For CaFeO$_2$ and BaFeO$_2$, the optimized crystal structures were used.}
    \begin{tabular}{ccccc}
      \hline
      \hline
      &$J_1$ & $J_2$ & $J_3$ & $J_4$\\
      \hline
      SrFeO$_2$ (expt.)&7.04 & 2.18 & 0.43 & -0.23 \\
      SrFeO$_2$ (opt.)&7.91 &  2.29 &  0.30 & -0.30 \\
      CaFeO$_2$ (opt.)&8.90 &  3.24 &  0.35 & -0.45 \\
      BaFeO$_2$ (opt.)& 5.81 &   1.34 & -0.29 & -0.16 \\
      \hline
      \hline
    \end{tabular}
    \label{table1}
  \end{table}


  \begin{figure}
    \includegraphics[width=6.5cm]{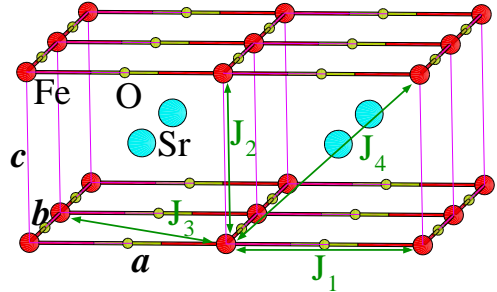}
    \caption{(color online) Perspective view of the tetragonal structure of SrFeO$_2$.
      The large, middle, and small spheres represent the Sr, Fe, and O
      ions, respectively.
      The spin exchange paths $J_1$, $J_2$, $J_3$, and $J_4$ are also indicated.}
    \label{fig1}
  \end{figure}

  \begin{figure}
    \includegraphics[width=8.0cm]{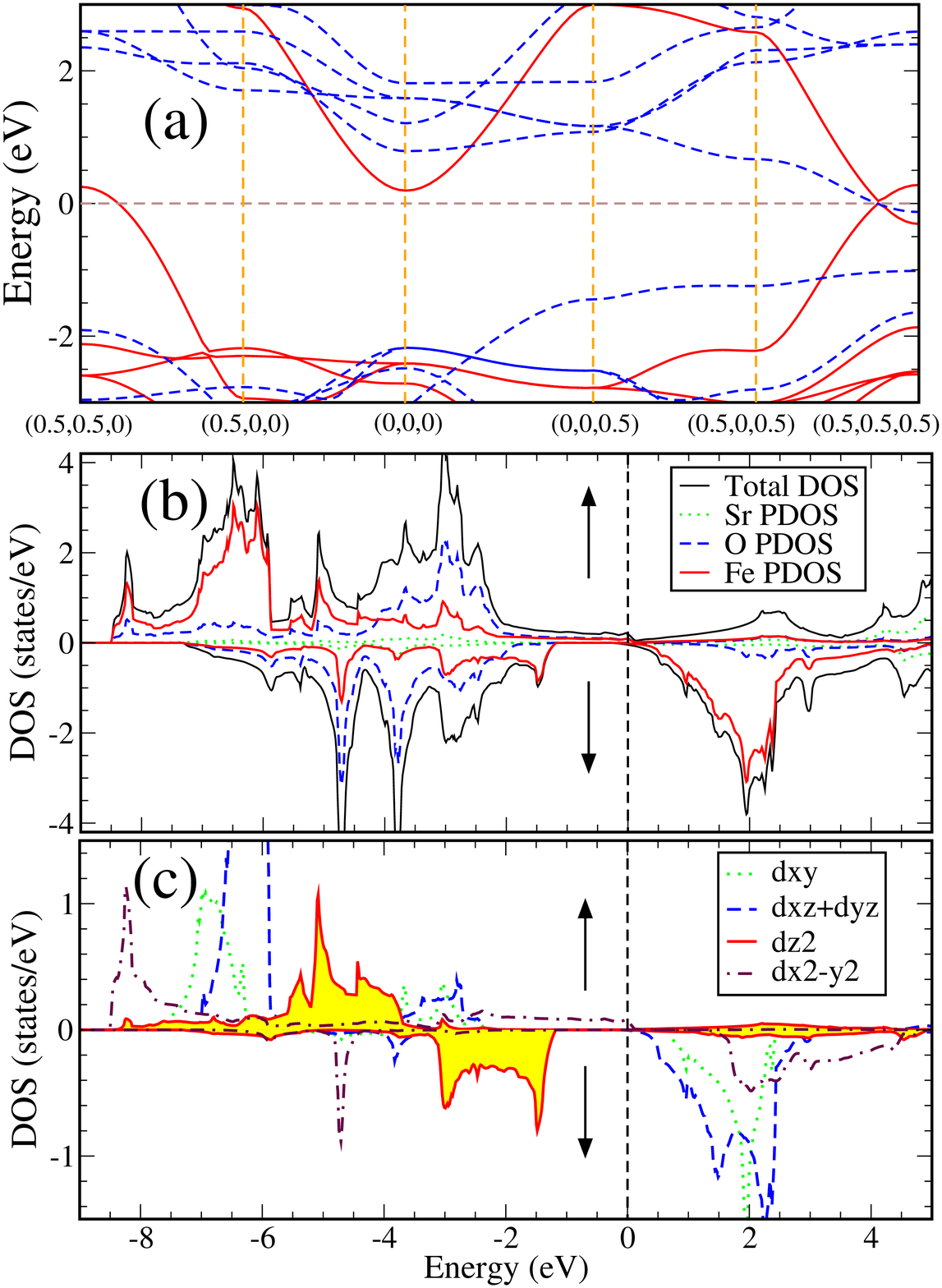}
    \caption{(color online) Electronic structure calculated for the FM
      state of SrFeO$_2$: (a) Dispersion relations of the up-spin and
      down-spin bands (solid and dashed lines, respectively) in the
      vicinity of the Fermi level. (b) Total DOS and PDOS plots for the Sr, Fe and O atom
      contributions. (c) PDOS plots for the Fe 3$d$ orbitals, where the PDOS for the
      Fe $d_{z^2}$ orbital was highlighted by shading.}
    \label{fig2}
  \end{figure}

  To probe the causes for these 
  apparently anomalous structural and magnetic
  properties
  in SrFeO$_{2}$, we examined the magnetic properties of SrFeO$_{2}$ by
  performing
  density functional theory (DFT) band structure and total energy
  calculations to
  evaluate
  its spin exchange interactions and performing Monte Carlo (MC)
  simulations
  to calculate the N\'eel temperature using the extracted spin
  exchange parameters.
  We show that 
  the down-spin Fe 3$d$ electron in SrFeO$_{2}$ occupies the nondegenerate $d_{z^2}$
  level rather than
  the 
  degenerate ($d_{xz}$, $d_{yz}$) levels, which explains
  the absence of
  Jahn-Teller instability, 
  the occurrence of
  the easy $ab$-plane magnetic anisotropy  
  as well as
  the strongly AFM coupling between the inter-layer and
  intra-layer
  nearest-neighbor (NN) Fe$^{2+}$ ions 
  thereby giving rise to the observed
  3D
  (0.5, 0.5, 0.5) AFM order in SrFeO$_{2}$. MC
  simulations of
  specific heat show that the strong inter-layer spin exchange is
  essential for the
  high N\'eel temperature. 
  Similar results are also found for the
  hypothetical isostructural
  compounds CaFeO$_{2}$ and BaFeO$_{2}$.

  Our first-principles spin-polarized DFT calculations for MFeO$_{2}$
  (M = Ca, Sr, Ba) were performed on the basis of the projector
  augmented wave method \cite{PAW} encoded in the Vienna ab initio
  simulation package \cite{VASP} using the local density approximation
  \cite{LDA} and the plane-wave cutoff energy of 400 eV.
  To properly describe the strong electron
  correlation associated with the Fe 3$d$ states, the LDA plus on-site
  repulsion U method (LDA$+$U) was employed \cite{Liechtenstein1995}.
  In the following, we report only those results
  obtained with $U=4.6$ eV and $J=0$ eV on Fe \cite{Xiang2007}, but
  the use of other U values between $3-6$ leads to qualitatively the
  same results.

  SrFeO$_2$ adopts the P4/mmm space group with $a=3.985$ \AA\ and
  $c=3.458$ \AA\ at 10 K \cite{Tsujimoto2007}. As shown in
  Fig.~\ref{fig1}, the FeO$_2$ layers separated by Sr$^{2+}$ ions are
  stacked along the $c$ axis. As a first step to discuss the magnetic
  properties of SrFeO$_2$, the electronic structure of SrFeO$_2$
  calculated for its ferromagnetic (FM) state
  [$\mathbf{q}=(0.0,0.0,0.0)$] is presented in Fig.~\ref{fig2}.
  The band dispersion relations of Fig.~\ref{fig2}(a) show that the FM
  state is metallic because the top portion of the up-spin valence
  bands (an O p and Fe d state) is above the bottom portion of the
  up-spin
  conduction bands (a Fe p and Sr d state) at $R$ (0.5, 0.5, 0.5) point.
  The plots of
  the density of states (DOS) and the partial DOS (PDOS) in
  Fig.~\ref{fig2}(b) reveal that there is a strong hybridization
  between the O 2$p$ and Fe 3$d$ states in the valence bands.
  Further analysis shows that the largest coupling occur at R-point between Fe
  $d_{x^2-y^2}$ orbital and O $p_x,p_y$ orbitals.
  When the PDOS plots of the Fe and O atoms in
  Fig.~\ref{fig2}(b) are compared with those of the Fe 3$d$
  states in Fig.~\ref{fig2}(c), it is seen that the up-spin bands have
  Fe 3$d$ energy levels below the O 2$p$ state, whereas the down-spin Fe
  $3d$
  bands is above. This reflects the large exchange splitting of the Fe
  3$d$
  levels, which is responsible for the high-spin state of the Fe$^{2+}$
  ion. 
  It is important to note from Fig.~\ref{fig2} that the occupied
  down-spin state has the $d_{z^2}$ character, not the doubly-degenerate
  ($d_{xz}$, $d_{yz}$) character as expected from crystal field theory
  for the $D_{4h}$ point symmetry \cite{Wells1962,Tsujimoto2007}.
  This is because in the layered structure, the energy of
  the $d_{z^2}$ state is significantly decreased due to the reduced Coloumb repulsion.
  Because the $d_{z^2}$ is
  non-degenerate, no Jahn-Teller type distortion
  is expected
  for SrFeO$_2$, as found experimentally \cite{Tsujimoto2007}.

\begin{figure}
    \includegraphics[width=7.0cm]{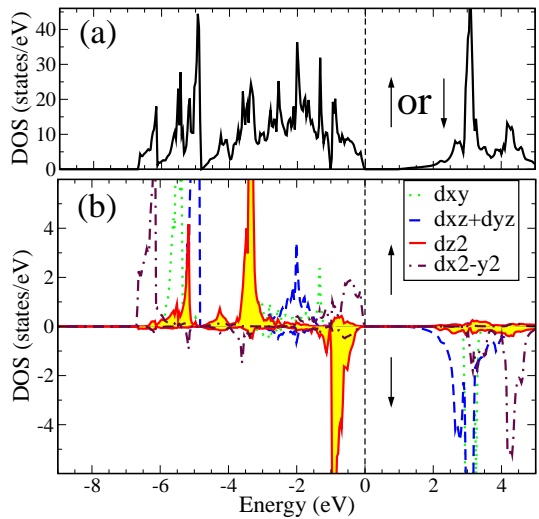}
    \caption{(color online) Electronic structure calculated for the AF1 state of SrFeO$_2$:
      (a) Total DOS plot, which is identical for the up-spin and down-spin bands.
      (b) PDOS plots calculated for the Fe 3$d$ orbitals, where the shaded
      regions refer to the Fe 3$d_{z^2}$ orbital contributions.}
    \label{fig3}
  \end{figure}

  To determine the magnetic ground state and discuss the magnetic
  properties of SrFeO$_2$, we considered four more ordered spin states
  besides the FM state, namely, the AF1 state with
  $\mathbf{q}=(0.5,0.5,0.5)$, the AF2 state with
  $\mathbf{q}=(0.0,0.0,0.5)$, the AF3 state with
  $\mathbf{q}=(0.5,0.5,0.0)$, and the AF4 state with
  $\mathbf{q}=(0.5,0.0,0.5)$. The experimentally observed AFM
  structure is AF1.  Our LDA+U calculations show that
  all the AFM states are lower in energy than the FM state, and the
  AF1 state is the ground state, in good agreement with experiment \cite{Tsujimoto2007}.
  These results are
  consistent with the facts that in the high-spin ($d^6$) configuration,
  there is no partially occupied $d$ state to stabilize the FM
  phase \cite{Dalpian2006}.
  The total DOS and PDOS plots
  presented in Fig.~\ref{fig3} show that SrFeO$_2$ in the AF1 phase has a
  band gap as expected for this magnetic semiconductor.
  Comparing to the electronic band structures of the FM phase,
  the band structure of the AF1 phase exhibit an important difference;
  overall, the $d$ bands are narrower in the AF1 than in the FM
  state.
  In particular, the up-spin $d_{x^2-y^2}$ band has a
  significantly narrower bandwidth in the AF1 state.
  This observation is readily accounted for by considering a
  spin-1/2 square-net lattice with one magnetic orbital per site,
  the hopping integral $t$ between adjacent sites and the on-site
  repulsion $U$. For the FM arrangement of the spins, the up-spin
  (or down-spin) states of adjacent sites are identical in energy and
  the interaction energy between them is $t$, so the widths of the
  up-spin and down-spin bands are proportional to $t$. For the AFM
  arrangement of spins, the up-spin (or down-spin) states of adjacent
  sites differ in energy by $U$ and the interaction energy between them
  is $t^2/U$, so the widths of the up-spin and down-spin
  bands are proportional to $t^2/U$. Since $t \gg t^2/U$ for usual
  magnetic
  solids, the width of the electronic energy band is
  much wider for the FM state than for the AFM state.

  To extract the values of the four spin exchange parameters $J_1$,
  $J_2$, $J_3$ and $J_4$ (see Fig.~\ref{fig1}), we map the relative
  energies of the five ordered spin states (i.e., FM, AF1, AF2, AF3
  and AF4) obtained from the LDA$+$U calculations onto the corresponding
  energies given by the Heisenberg spin Hamiltonian made up of the
  four spin exchange parameters. 
  This mapping analysis was carried out as described in ref. \cite{mapping}.  
  The obtained exchange parameters
  are summarized in Table.~\ref{table1}. The
  intra-layer NN spin exchange $J_1$ is quite
  strong, which is due to the strong 180$^{\circ}$ Fe-O-Fe
  superexchange between the $d_{x^2-y^2}$ orbitals mediated by O $p_x$
  and $p_y$  orbitals \cite{Kanamori1958,Goodenough1955}. The inter-layer NN
  spin exchange $J_2$ is strongly AFM and is weaker than $J_1$ only by
  a factor of $\sim 3$. The inter-layer spin exchange $J_2$ originates from
  the direct through-space overlap between the $d_{xz}$/$d_{yz}$
  orbitals of Fe.

  \begin{figure}
    \includegraphics[width=7.0cm]{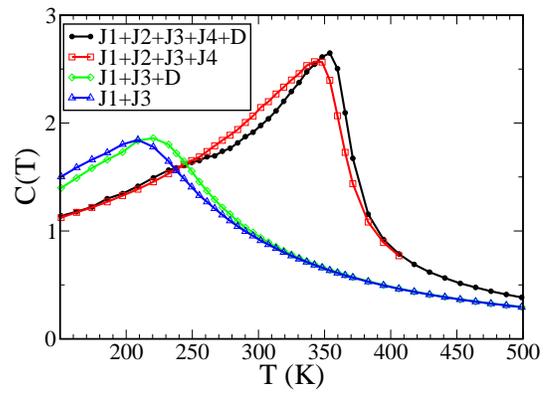}
    \caption{(color online) Specific heat of SrFeO$_2$, $C$, calculated as
      a function of temperature $T$ on the basis of the classical spin Hamiltonian
      defined in terms of the spin exchange and the spin anisotropy parameters.}
    \label{fig4}
  \end{figure}

  To account for the observed anisotropic magnetic property of
  SrFeO$_2$, we carried out LDA+U calculations for the FM state by
  including spin-orbit coupling (SOC) interactions. These LDA+U+SOC
  calculations show that 
  the state with the spin moments parallel to the $ab$-plane
  is more stable than the state with the spin moments parallel to the
  $c$-axis by $4$ meV/Fe, while 
  the calculated orbital moment is 0.22
  $\mu_B$ for the $\perp c$ spin arrangement, and 0.01 $\mu_B$ for the
  $\parallel c$ spin arrangement. 
  This easy $ab$-plane anisotropy, which is in accord
  with experiment \cite{Tsujimoto2007},
  can be
  explained by analyzing the SOC Hamiltonian \cite{Xiang2007B}. Since
  the up-spin and down-spin $d$ bands are well separated due to
  the large exchange splitting, one can neglect interactions between
  the up-spin and down-spin states under the SOC. With $\theta$  and
  $\phi$ as the zenith and azimuth angles of the magnetization in the
  direction $\mathbf{n}$($\theta$, $\phi$),
  the
  spin-conserving
  term of
  the $\lambda \hat{\mathbf {L}} \cdot \hat{\mathbf {S}}$ operator
  ($\lambda < 0$ for Fe$^{2+}$) is given by \cite{Wang1996,Xiang2007B}
  \begin{equation}
    \lambda \hat{S}_n (\hat{L}_z \cos \theta + \frac{1}{2} \hat{L}_+ e^{-i \phi}
    \sin \theta  +\frac{1}{2}
    \hat{L}_- e^{i \phi} \sin \theta )
  \end{equation}
  If SrFeO$_2$ were to have one down-spin electron in the
  doubly-degenerate level ($d_{xz},d_{yz}$), it should have uniaxial
  magnetic properties with the spin moments parallel to the $c$-axis
  ($\theta=0^{\circ}$) according to the degenerate perturbation theory
  \cite{Dai2005}. Our calculations show that, for the down-spin part, the
  highest occupied Fe $3d$ level is $d_{z^2}$ while the lowest
  unoccupied Fe $3d$ level is ($d_{xz},d_{yz}$). When the spin lies in
  the $ab$-plane ($\theta=90^{\circ}$), the mixing between $d_{z^2}$ and
  $d_{xz},d_{yz}$ due to the raising and lowering operators is the largest. Thus, nondegenerate
  perturbation theory shows that SrFeO$_2$ has an easy $ab$- plane
  anisotropy with a relative large orbital moment.

  To explain why the layered compound SrFeO$_2$ has a very high
  N\'eel
  temperature (473 K),
  we perform MC simulations for a $12 \times 12
  \times 12$ supercell based on the classical spin Hamiltonian:
  \begin{equation}
    H= \sum_{<ij>} J_{ij} \vec{S_i} \cdot \vec{S_j} + \sum_{i} D
    S_{iz}^2,
  \end{equation}
  where the spin exchange parameters $J_{ij}$ are those defined in
  Fig.~\ref{fig1}, $D= 1$ meV (i.e., $DS_{iz}^2 = 4 D = 4$ meV) is the
  spin anisotropy parameter, and $S=2$.
  To obtain $T_N$, we first
  calculate the specific heat
  $C=(\langle E^2 \rangle-\langle E \rangle^2)/T^2$
  after the system reaches equilibrium at a given temperature ($T$). 
  Then $T_N$ can be
  obtained
  by locating the peak position in the $C(T)$
  vs. $T$
  plot, shown in Fig.~\ref{fig4}.
  For SrFeO$_2$ the calculated $T_N$ is 354 K, which is in reasonable
  agreement with the experimental value of 473 K. We should note that
  a smaller $U$ value leads to larger exchange parameters $J_1$ since
  $J_1 \propto t^2/U$, and
  $J_2$ changes a little, thus $T_N$ will be higher.
  For instance, if $U=4.0$,
  then $J_1=8.68$ meV and $J_2=2.23$ meV, resulting in $T_N=453$ K.

  It was shown that large
  magnetic anisotropy energy can stabilize long-range magnetic 
 order in a one-dimensional (1D) monatomic Co chains \cite{Gambardella2002}.
  To determine how $T_N$  
  is affected by the different spin exchange parameters
  and the spin anisotropy, we performed three additional simulations;
  one with
  the
  spin anisotropy $D$ neglected, one with
  the
  inter-layer exchange parameters ($J_2$ and $J_4$) neglected, and one
  with both the inter-layer spin exchange parameters ($J_2$ and $J_4$)
  and the spin anisotropy $D$ neglected. The resulting $C(T)$ vs. $T$
  plots are presented in Fig.~\ref{fig4}. As can be seen, when $D$ is
  disregarded, $T_N$ is only slightly lower (347 K). If only the
  inter-layer exchange parameters are neglected, the $C(T)$ vs. $T$
  plot shows a broad peak at 220 K. A broad peak occurs at 208 K if
  both the inter-layer exchange parameters and the spin anisotropy are
  neglected. At any nonzero temperature, a 1D or 2D isotropic
  Heisenberg model with finite-range exchange interaction can be
  neither FM nor AFM \cite{Mermin1966}.
  Thus, the
  broad peak in the $C(T)$ vs. $T$ plot, obtained when the inter-layer
  spin exchange 
  interactions are neglected, indicates the presence of short-range
  order. Our MC simulations indicate that the inter-layer interactions
  are primarily responsible for the high $T_N$ of SrFeO$_2$.

  It should be noted that CaFeO$_{3-x}$ and BaFeO$_{3-x}$ have
  structures and properties
  similar
  to those of SrFeO$_{3-x}$. Thus, it is of
  interest to consider the structural and magnetic properties of
  hypothetical CaFeO$_2$ and BaFeO$_2$ assuming that they are
  isostructural with SrFeO$_2$. For this purpose, the structures of
  MFeO$_2$ (M = Sr, Ca, Ba) were optimized by performing LDA+U
  calculations for the AF1 state. The optimized lattice constants $a$
  and $c$ for SrFeO$_2$ are 3.92 \AA\ and 3.40 \AA, respectively,
  which are in good agreement with experiment.
  The
  calculated lattice constants are $a=3.86$ \AA\ and $c=3.12$ \AA\
  for CaFeO$_2$, and $a=3.98$ \AA\ and $c=3.79$ \AA\ for BaFeO$_2$. The
  trend in the lattice constants is consistent with the ionic radii of
  Ca$^{2+}$, Sr$^{2+}$, and Ba$^{2+}$. The exchange parameters 
  calculated for the optimized 
  MFeO$_2$ (M = Sr, Ca, Ba) structures are 
  listed in
  Table.~\ref{table1}. 
  All
  these compounds should have the AF1 state as the ground state since
  the spin exchange interactions are dominated by AFM $J_1$ and $J_2$.
  Among MFeO$_2$ (M = Ca, Sr, Ba), CaFeO$_2$ has the largest $J_1$ and
  $J_2$ values, while BaFeO$_2$ has the smallest $J_1$ and $J_2$
  values. This trend reflects the fact that the strengths of these
  interactions increase with decreasing the lattice constants. Thus, CaFeO$_2$
  is predicted to have a higher $T_N$
  than does SrFeO$_2$.

  In summary, SrFeO$_2$ has no Jahn-Teller instability because the occupied 
  down-spin d-level is $d_{z^2}$. The N\'eel temperature of SrFeO$_2$ is high because 
  the intra-layer NN spin exchange is strong while the inter-layer NN spin 
  exchange is substantial. The in-plane magnetic anisotropy of SrFeO$_2$ arises 
  from the SOC-induced interaction between the $d_{z^2}$ and
  ($d_{xz}$, $d_{yz}$) states.

  Work at NREL was supported by the U.S. Department of
  Energy, under Contract No. DE-AC36-99GO10337.
  The research at NCSU was supported by the Office of Basic
  Energy Sciences, Division of Materials Sciences, U.S.
  Department of Energy, under Grant No. DE-FG02-86ER45259.




\begin{thebibliography}{99}

  \bibitem{Shao2004}Z. Shao  and S. M. Haile, Nature {\bf 431}, 170
    (2004).

  \bibitem{Falcon2002}H. Falc\'on {\it et al.}, Chem. Mater. {\bf 14}, 2325 (2002).

  \bibitem{Badwal2001}S. P. S. Badwal and F. T. Ciacchi,
    Adv. Mater. {\bf 13}, 993 (2001)

  \bibitem{Wang2001}Y. Wang {\it et al.},  Mater. Lett. {\bf 49}, 361 (2001).

  \bibitem{Lebon2004}A. Lebon {\it et al.},
    Phys. Rev. Lett. {\bf 92}, 037202 (2004).

  \bibitem{Tsujimoto2007} Y. Tsujimoto {\it et al.}, Nature {\bf 450}, 1062
    (2007).

  \bibitem{Hayward2007}M. A. Hayward and M. J. Rosseinsky,  Nature {\bf 450}, 960 (2007).

  \bibitem{Wells1962}A. F. Wells,  {\it Structural Inorganic Chemistry} 3rd edn (Oxford
    Univ. Press, Oxford, UK, 1962).

  \bibitem{Murakami1998}Y. Murakami {\it et al.},
    Phys. Rev. Lett. {\bf 81}, 582 (1998).

  \bibitem{Dai2005}D. Dai and M.-H. Whangbo, Inorg. Chem. {\bf 44}, 4407 (2005).

  \bibitem{PAW}P. E. Bl\"ochl, Phys. Rev. B {\bf 50}, 17953 (1994);
    G. Kresse and D. Joubert, {\it ibid}  {\bf 59}, 1758 (1999).

  \bibitem{VASP}G. Kresse and J. Furthm\"uller, Comput. Mater. Sci. {\bf
    6}, 15 (1996); Phys. Rev. B {\bf 54}, 11169 (1996).

  \bibitem{LDA}J. P. Perdew and A. Zunger, Phys. Rev. B {\bf 23}, 5048 (1981);
    D. M. Ceperley and B. J. Alder, Phys. Rev. Lett. {\bf 45}, 566 (1980).

  \bibitem{Liechtenstein1995}A. I. Liechtenstein {\it et al.}, Phys. Rev. B  {\bf 52},
    R5467 (1995).

  \bibitem{Xiang2007}H. J. Xiang and M. -H. Whangbo,
    Phys. Rev. Lett. {\bf 98}, 246403 (2007).

  \bibitem{Dalpian2006} G. M. Dalpian {\it et al.},
    Solid State Commun. {\bf 138}, 353 (2006).
  
  \bibitem{mapping}H. J. Xiang, C. Lee, and M. -H. Whangbo,
    Phys. Rev. B {\bf 76}, 220411(R)(2007). 
    
  \bibitem{Kanamori1958}J. Kanamori, J. Phys. Chem. Solids {\bf 10}, 87 (1958).

  \bibitem{Goodenough1955} J. B. Goodenough, Phys. Rev. {\bf 100}, 564 (1955).


  \bibitem{Xiang2007B}H. J. Xiang and M. -H. Whangbo, Phys. Rev. B  {\bf
    75}, 052407 (2007).

  \bibitem{Wang1996}X. Wang {\it et al.},
    Phys. Rev. B {\bf 54}, 61 (1996).

  \bibitem{Gambardella2002}P. Gambardella {\it et al.}, Nature {\bf 416}, 301 (2002).

  \bibitem{Mermin1966}N. D. Mermin  and H. Wagner, Phys. Rev. Lett. {\bf
    17}, 1133 (1966).



  \end{thebibliography}
\end{document}